\documentclass[prl,twocolumn,showpacs,amsmath,amssymb]{revtex4}
\usepackage{graphicx}
\usepackage{dcolumn}
\usepackage{bm}
\usepackage{bbm}

\newcommand{\ket}[1]{\mbox{$ | #1 \rangle $}}
\newcommand{\bra}[1]{\mbox{$ \langle #1 | $}}
\newcommand{\braket}[1]{\ensuremath{\left\langle #1\right\rangle}}

\begin{document}

\title{Squashing Models for Optical Measurements in Quantum Communication}
\author{Normand J. \surname{Beaudry}$^1$, Tobias \surname{Moroder}$^{1,2}$, Norbert \surname{L\"utkenhaus}$^{1,2}$} \affiliation{$^1$ Institute for Quantum Computing, University of Waterloo,   Waterloo, Canada \\ $^2$ Quantum Information Theory Group, Institute of Theoretical Physics I, and  Max-Planck Research Group, Institute of Optics, Photonics and Information, University Erlangen-Nuremberg, Erlangen, Germany }

\date{\today}

\date{\today}

\begin{abstract}
Measurements with photodetectors are naturally described in
the infinite dimensional Fock space of one or several modes. For some measurements a model has been postulated which describes the full mode measurement as a composition of a mapping (squashing) of the signal into a small dimensional Hilbert space followed by a
specified target measurement. We present a formalism to investigate whether a
given measurement pair of full and target measurements can be connected by a squashing model. We show that a measurement used in the BB84
protocol does allow a squashing description, although the corresponding six-state
protocol measurement does not. As a result, security proofs for the BB84 protocol can be based on the assumption that the eavesdropper forwards at most one photon, while the same does not hold for the six-state protocol.
\end{abstract}

\pacs{42.50.Ex, 03.67Dd, 03.67.Hk, 42.79.Sz}
\maketitle

Detection devices play an important role in quantum communication protocols. In the theoretic design of these protocols, signals are often thought of as qubits, and therefore low-dimensional Hilbert spaces only need to be considered. In optical implementations, the signals are realized by photons, which are naturally described by the Fock spaces of spatio-temporal modes. Our goal is to determine how one can reduce the large-dimensional description of optical measurements of these modes to a particular lower-dimensional one. Our insight will provide a powerful tool to ease the analysis of optical implementations of quantum communication protocols.

A typical measurement in quantum communication is the one used in the BB84 QKD protocol \cite{bennett84a}, in which the incoming
light is split by a polarizing beam-splitter, which can be oriented either along the horizontal/vertical basis (labelled as $z$) or in the
+45/-45 degree basis (labelled as $x$). The signal is then sent to a threshold detector which cannot resolve the number of photons by which they are triggered. This measurement can be described as a single Positive Operator Valued Measure (POVM) with non-commuting POVM elements if the basis choice is done at random with some fixed probabilities. It has been postulated that there exists a squashing model for this set-up, which first maps (squashes) the incoming signal to a one-photon polarization Hilbert space, followed by the same BB84 measurement. A recent important security proof \cite{ma07a} is based on this detector property.

In this Letter, we define a squashing model and lay out a framework to determine whether a
given detection device allows a squashing model. We then
prove for the BB84 measurement that a squashing model
exists. Surprisingly, the corresponding measurement in the
six-state protocol \cite{bruss98a,bechmann99a} does not admit a squashing model. More details of these results will be presented in a future paper.

\begin{figure}
 \includegraphics[scale=0.38]{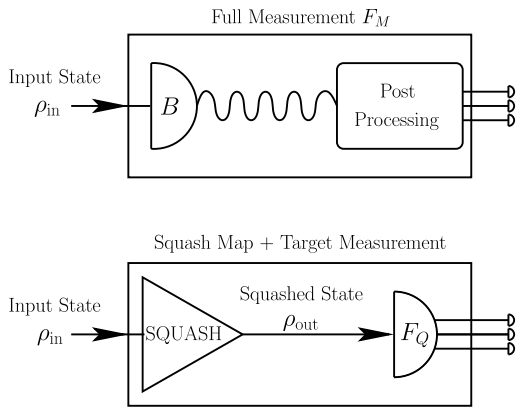}
\caption{The full measurement $F_{M}$ (above) has a general optical input $\rho_{in}$, which is first measured by a receiver's physical detector $B$, followed by classical post-processing. The squashed measurement (below) has the same general optical input $\rho_{in}$, which is then squashed by a map $\Lambda$ to a smaller Hilbert space, followed by a fixed physical measurement $F_{Q}$. It is required that both of these measurements produce the same output statistics for all $\rho_{in}$.}
\label{fig:squashdefn}
\end{figure}

First, we will define a squashing model more precisely. A full measurement, $F_M$, described by a POVM with elements $F_M^{(i)}$
defined on a large (possibly infinite dimensional) Hilbert space $M$ is said to {\em admit a squashing model with
respect to a target measurement, $F_Q$,} with POVM elements $F_Q^{(i)}$ on a
smaller dimensional Hilbert space $Q$ if a squashing map
$\Lambda$ from $M$ to $Q$ exists, such that the composition of the
squashing map and the measurement on $Q$ is statistically equivalent to the
measurement on system $M$.  In other words, the two measurement
models in Fig.~\ref{fig:squashdefn} must act identically for any
input signal. 

The measurement description via the POVM elements $F_M^{(i)}$ and
$F_Q^{(i)}$ need not correspond to the basic events by the
detectors, such as the pattern of detector clicks, but can involve some post-processing.  For example, in the
optical implementation of the BB84 measurement above, double clicks occur if both detectors fire due to a multi-photon input, while after squashing, at most one
photon is contained in the signal and so no double clicks can occur. Therefore, to
match the number of possible outcomes, we can choose to map
double clicks of the full measurement randomly to the
single-click event of one of the two detectors. This mapping has been
introduced before in the security analysis of QKD \cite{nl99a,nl00a}. 

In the context of QKD, one typically assumes the \emph{calibrated device scenario} in which the detection device is trusted and known. Then if a squash model exists, the corresponding squashing map can become part of the eavesdropper's (Eve's) attack. Therefore we can assume, without loss of generality, that Eve sends a signal in the Hilbert space $Q$
to the receiver, Bob. As an example, many security proofs assume that Eve forwards polarized single photons (qubits) or vacuum states to the receiver. If a given full optical implementation of a polarization measurement has a squash model connecting it to the single photon polarization measurement assumed in the security proof, then this proof is also valid for the full optical implementation of the protocol. Additionally, squashing the detection to a finite-dimensional system makes it possible to use the fast converging de Finetti theorems of Renner \cite{renner07a} on the level of the squashed system, even if the original full system is infinite dimensional.

Notice that the existence of a squashing model for a given full measurement $F_M$ and target measurement $F_Q$ is
the question of the existence of a particular squasher connecting these measurements. Any valid
squasher must be a trace preserving completely positive map, $\Lambda$,  and can be described by a set of Kraus operators $\{A_{k}\}$,
which obey $\sum_{k}A_{k}^{\dag}A_{k}=\mathbbm{1}_{M}$. The statistical equivalence of the full measurement $F_{M}$ and concatenation of $\Lambda$ and $F_{Q}$ can be stated formally as
\begin{equation}
\begin{split}
 \!\!\!\!\mathrm{Tr}\left[\rho_{in} F_{M}^{(i)}\right]
 &=\mathrm{Tr}\left[\!\Lambda(\rho_{in})F_{Q}^{(i)}\!\right]\!=\!\mathrm{Tr}\!\left[\!\sum_{k}\!A_{k}\rho_{in}
 A_{k}^{\dag}F_{Q}^{(i)}\!\right]\\ 
=\mathrm{Tr}&\left[\!\rho_{in}\sum_{k}\!A_{k}^{\dag}F_{Q}^{(i)}A_{k}\!\right]\!=\!\mathrm{Tr}\left[\rho_{in}
\Lambda^{\dag}(\!F_{Q}^{(i)})\!\right] 
\end{split}
\label{eq:trace}
\end{equation}
where $\rho_{in}$ is the density matrix of the incoming signal. We require
Eqn.~(\ref{eq:trace}) to hold for all incoming signals $\rho_{in}$, which is
fulfilled if and only if  
\begin{equation}
 F_{M}^{(i)}=\Lambda^{\dag}(F_{Q}^{(i)})=\sum_{k}A_{k}^{\dag}F_{Q}^{(i)}A_{k}
\label{eq:POVMmap}
\end{equation}
holds. That is, the  adjoint squashing map $\Lambda^\dag$ with Kraus operators
$A_{k}^{\dag}$ map each qubit POVM operator to the corresponding
POVM operator for the mode detector. The adjoint map is again a
completely positive map. It is not necessarily trace
preserving, but it is unital.

The question for the existence of a suitable adjoint squashing map
$\Lambda^{\dag}$ has been formulated as the search for a suitable
set of Kraus operators $\{A_k^{\dag}\}$. As the Kraus operators are not unique, we
reformulate the condition Eqn.~(\ref{eq:POVMmap}) using the
Choi-Jamio\l kowski isomorphism \cite{jamiolkowski72a,bengtsson}. It relates the
map $\Lambda^\dag$ to a bipartite operator $\tau$ on a duplicated
output Hilbert space $Q Q'$ by applying the map to half of a
maximally entangled state $\ket{\psi^+}=1/\sqrt{d}\sum_{i=1}^{d}|i\rangle_{Q}|i\rangle_{Q'}$, where $d=\mathrm{dim}(QQ')$, by $\tau = \Lambda^{\dag} \otimes \mathrm{id}\left(|\psi^+\rangle\langle\psi^+|\right)$. From this representation
  one can form the transfer matrix
  $\tau^R$ by reordering the coefficients via $\langle
  k,k'|\tau^R |l,l'\rangle = \langle k,l|\tau |k',l'\rangle$. Given an operator $O=\sum_{i,j}
  o_{i,j} |i\rangle \langle j|$, we introduce its vector notation as $|O\rangle\rangle
    = \sum_{i,j} o_{i,j} |i\rangle |j \rangle$, and so we can write
    $|\Lambda^\dag(O)\rangle\rangle = \tau^R
      |O\rangle\rangle$.
In this formulation, the search for a squashing model for a full measurement $F_M$ and a target measurement $F_Q$ is the search for a map $\tau$ such that
\begin{subequations}
 \begin{gather}
  \tau^R |F_{Q}^{(i)}\rangle\rangle = |F_{M}^{(i)}\rangle\rangle, \label{eq:restricta}\\
\langle k,k'|\tau^R |l,l'\rangle = \langle k,l|\tau |k',l'\rangle, \\
\tau^{\dag}=\tau\geq0.
 \end{gather}
\label{eq:restrict}
\end{subequations}
\noindent Here, $\tau$ corresponds to the adjoint map $\Lambda^{\dag}$. The constraint that $\Lambda^{\dag}$ be unital, and therefore $\Lambda$ trace-preserving, is already contained in the above conditions, as the POVM elements on $M$ and $Q$ each add up to the identity operator in their respective Hilbert spaces, as can be easily seen in the formulation of Eqn.~(\ref{eq:POVMmap}). Overall, we have reformulated the search for a suitable squashing operation as the search for a positive semidefinite operator $\tau$ that satisfies a fixed number of linear constraints, which can be efficiently solved using convex optimization. Searching for completely positive maps using these techniques has been used, for example, in \cite{reimpell05a,fletcher07a}.

\begin{figure}[ht]
 \includegraphics[scale=0.29]{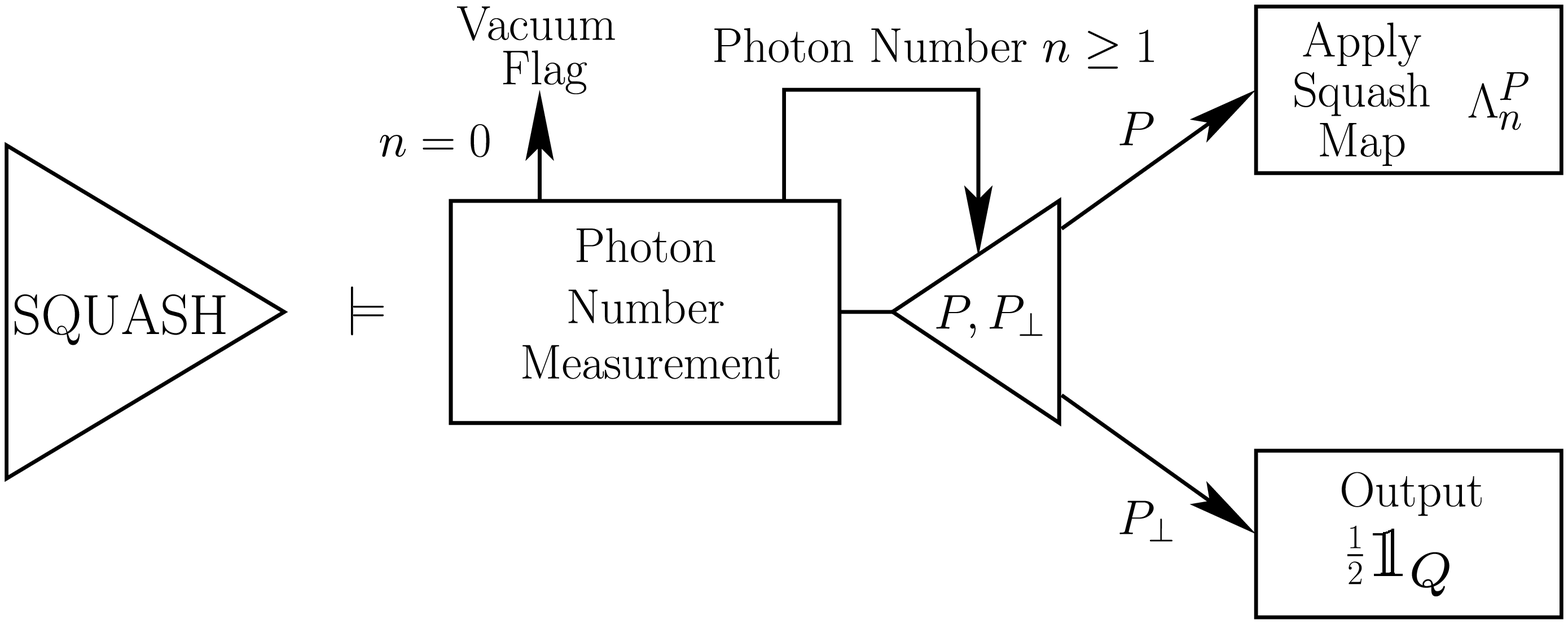}
 \caption{Reduction of the considered squashing operation of the
   BB84 protocol. The squashing operation can be modelled as a
   photon number measurement followed by a projection
   measurement onto a 4-dimensional subspace. Depending on the outcome of these measurements, one
   either proceeds with a  low-dimensional squashing operation $\Lambda_{n}^P$ or outputs a completely mixed qubit state.}
 \label{fig: red_squash}
\end{figure}

To simplify the search for the appropriate squashing operation, we can
exploit further properties of the physical measurement. Typical
measurement schemes only involve photon counting and hence commute with a quantum non-demolition (QND) measurement of the total number of photons. Consequently, we can decompose the squashing operation into a
photon number measurement, followed by the appropriate squashing
operation conditioned on the outcome of the QND measurement, as schematically
indicated in Fig.~\ref{fig: red_squash}. This model now casts the problem into finite dimensions, since we only
need to find the corresponding map for each finite dimensional photon number
subspace.

We now consider the situation where we choose as target measurements the full measurement restricted to the Fock space containing zero or one photon, which is a qutrit space. As the resulting POVM elements $F_{Q}^{(i)}$ still commute with a QND measurement of the total photon number, this means that the squashing map can be thought of as statistically outputting either no photon or one photon. We can now split off the zero-photon case easily in the typical scenario, where the full and target measurements have the vacuum projection as one POVM element, while none of the other elements contains a vacuum component. As a result, the squasher will output a vacuum signal if and only if the photon number $n$ measured in the QND measurement on the input mode space is zero. To simplify the presentation, we split these events off as a flag (see Fig.~\ref{fig: red_squash}) sent by the squasher, signalling that the input signal contains no photon, and we can now restrict ourselves to the case that for $n \geq 1$ input photons, the squasher outputs exactly one photon in the relevant modes, which enters the target measurement. In the case of the BB84 and the six-state measurements two polarization modes are sufficient to describe the multi-photon Hilbert space, so we can assume that for $n\neq 0$ exactly one qubit in the form of a photon with polarization degrees of freedom is output from the squashing operation. In this formulation, the POVM elements $F_Q^{(i)}$ are now restricted to the full measurements of the one-photon Hilbert space, as the vacuum events have been replaced by the flag structure of the squasher.

As a third step, we refine the squasher further by using the specific structure of the BB84 measurement. Here the
full measurement operators on the n-photon subspace ($n\geq 1$) can be
conveniently written as
\begin{equation}
\label{eq:measurementoperators}
  F_{M,\;n}^{(b,\alpha)}\!=\!\frac{(-1)^b}{4} \left(
    \ket{n,0}_{\alpha}\bra{n,0} - \ket{0,n}_{\alpha}\bra{0,n} \right)+
    \frac{\mathbbm{1}}{4},
\end{equation}
where $\alpha\in \{x,z\}$ labels the basis choice for the polarizing
beamsplitter, $b\in \{0,1\}$ corresponds to the ``0'' or ``1''
outcome of the detector, and $\ket{l,k}_{\alpha}$ is a two-mode Fock state with photon numbers $l$ and $k$ with respect to the polarization mode basis $\alpha$.
We define a subspace $P$  
spanned by the $4$ vectors $\ket{n,0}_{\alpha}$ and
$\ket{0,n}_{\alpha}$, and its orthogonal complement $P_\perp$ in the n-photon subspace. A QND measurement with respect to these two subspaces commutes with each
measurement POVM $F_Q^{(i)}$, and thus can precede the target detection 
scheme without loss of generality. We can therefore define independent squashing maps for each of the two sub-spaces, similarly to the treatment of the Fock spaces of photon number $n$. It is now easy to identify the squashing map starting on the $P_\perp$-subspace since the POVM elements $F_{M,n}^{(b,\alpha)}$ restricted to this subspace are given by $\mathbbm{1}_{P_\perp}/2$. An obvious choice for the squashing map here is to output the completely mixed qubit state, which triggers each POVM $F_{Q}^{(b,\alpha)}$ with equal probability (see Fig.~\ref{fig: red_squash}). This means we can now focus on the remaining part of the squashing operation, namely for all $n\geq 1$ the maps $\Lambda^P_n$ from the four-dimensional subspace $P$ of the n-photon Fock space to the qubit
space.

If the incoming signal is projected onto the subspace $P$, then either the map $\tau_{odd}$ or $\tau_{even}$ will be applied, depending on the parity of photon number $n$. First consider the case where $n\geq3$, the outcome of the QND measurement of the total photon number, is odd; the case $n=1$ is trivial. We use the following orthonormal basis to represent the 4-dimensional subspace $P$: $|\phi_{1}\rangle=|n,0\rangle_{z},|\phi_{2}\rangle=|0,n\rangle_{z}$, and
\begin{equation}
\begin{array}{c}
 |\phi_{3}\rangle=\frac{1}{C_{1}}\left(\sqrt{2^{n-2}}(|n,0\rangle_{x}+|0,n\rangle_{x})-|n,0\rangle_{z}\right)\\
 |\phi_{4}\rangle=\frac{1}{C_{1}}\left(\sqrt{2^{n-2}}(|n,0\rangle_{x}-|0,n\rangle_{x})-|0,n\rangle_{z}\right)
\end{array},
\label{eq:oddbasis}
\end{equation}
where we define $C_{g}\equiv\sqrt{2^{n-g}-1}$. The qubit measurement operators $F_{Q}^{(b,\alpha)}$ are given by:
\begin{equation}
\left\{\!
\left(\!
\begin{array}{c c}
 \frac{1}{2} & 0 \\
 0 & 0
\end{array}\!
\right)\!,
\left(\!
\begin{array}{c c}
 0 & 0 \\
 0 & \frac{1}{2}
\end{array}
\!\right)\!,
\frac{1}{4}\left(\!
\begin{array}{c c}
 1 & 1 \\
 1 & 1
\end{array}\!
\right)\!,
\frac{1}{4}\left(\!\!
\begin{array}{c c}
 1 & -1 \\
 -1 & 1
\end{array}
\!\!\right)\!
\right\}
\end{equation}
in the standard basis.  The full measurement operators $F_{M,\;n}^{(b,\alpha)}$ from Eqn.~(\ref{eq:measurementoperators}) in the basis given by Eqn.~(\ref{eq:oddbasis}) are
\begin{equation*}
 \begin{split}
  F_{M,\;n}^{(b,z)}\!=\!\left[\!\!\begin{array}{c c c c}
                   \frac{1-b}{2} & 0 & 0 & 0 \\
                   0 & \frac{b}{2} & 0 & 0 \\
                   0 & 0 & \frac{1}{4} & 0 \\
                   0 & 0 & 0 & \frac{1}{4}
                  \end{array}\!\!\right]\!,
  F_{M,\;n}^{(b,x)}\!=\!\frac{\mathbbm{1}}{4}+\frac{(-1)^{b}}{4}\left[\!\!\begin{array}{c c c c}
                   0 & s & 0 & t \\
                   s & 0 & t & 0 \\
                   0 & t & 0 & u \\
                   t & 0 & u & 0
                  \end{array}\!\!\right]
 \end{split}
\end{equation*}
where $\mathbbm{1}$ is the $4\times4$ identity matrix and we define the constants $s\equiv2^{1-n}, t\equiv s C_{1}, u\equiv1-s$. To obtain their vectorized form $|F_{Q}^{(b,\alpha)}\rangle\rangle$ and $|F_{M}^{(b,\alpha)}\rangle\rangle$, one needs to concatenate the columns of their matrix form into vectors.

Now we are ready to impose Eqs.~(\ref{eq:restrict}) on the adjoint squashing map. First note that $\tau^{R}$ maps real vectors into real vectors (Eqn. (\ref{eq:restricta})), and therefore the complex conjugate $(\tau^{R})^{*}$ also maps these vectors to each other. As a result, the average of these two also performs the mapping, and so we can assume that $\tau^R$ is a matrix with real entries. Also, the target measurement operators, $|F_{Q}^{(b,\alpha)}\rangle\rangle$, only span a three dimensional vector-space, so the matrix $\tau^R$ is not completely determined by the linear constraints. Keeping the undetermined entries as open parameters $a_i$, we then obtain $\tau_{odd}$, which is given by
\begin{equation*}
\left[
\begin{array}{c c c c c c c c}
 1 & 0 & 0 & a_{1} & 0 & a_{2} & 0 & a_{3} \\
 0 & 0 & s-a_{1} & 0 & -a_{2} & 0 & t-a_{3} & 0 \\
 0 & s-a_{1} & 0 & 0 & 0 & a_{4} & 0 & a_{5} \\
 a_{1} & 0 & 0 & 1 & t-a_{4} & 0 & -a_{5} & 0 \\
 0 & -a_{2} & 0 & t-a_{4} & \frac{1}{2} & 0 & 0 & a_{6} \\
 a_{2} & 0 & a_{4} & 0 & 0 & \frac{1}{2} & u-a_{6} & 0 \\
 0 & t-a_{3} & 0 & -a_{5} & 0 & u-a_{6} & \frac{1}{2} & 0 \\
 a_{3} & 0 & a_{5} & 0 & a_{6} & 0 & 0 & \frac{1}{2}
\end{array}
\right].
\end{equation*}
Using the assignment $a_{1}=s,a_{2}=0,a_{3}=t,a_{4}=0,a_{5}=0,a_{6}=1/2-s$ for the open parameters ensures that $\tau$ is positive semidefinite. By considering suitable subdeterminants it can be shown that these parameters must be chosen this way, and therefore the squashing map is unique. Further details will be included in a future paper. Following a similar procedure we can also construct the unique adjoint squashing operation for even $n\geq2$. Therefore, the squashing operation for the BB84 detector with active basis choice and the described post-processing exists.

The six-state protocol adds another measurement
direction to the BB84 setting, which sorts the polarization of the incoming
photons according to a circular basis choice (labelled $y$). Using the same
post-processing scheme of the double clicks results in similar
measurement operators as given by Eqn.~(\ref{eq:measurementoperators}) with $\alpha\in \{x,y,z\}$ as well as performing a renormalization. Hence the overall measurement description of 
the six-state protocol is similar to the BB84 case, where the transfer matrix $\tau^{R}$ is now completely determined by the linear constraints, as the POVM elements of $F_{Q}$ span the whole operator space. However,
this measurement device cannot be squashed down to the qubit level, since $\tau\not\geq 0$.
We can verify this statement independent of any of the reductions introduced earlier: all we need to show is that $\tau=\Lambda^{\dag}\otimes\mathrm{id}(\ket{\psi^+}\bra{\psi^+}) \not
\geq 0$. Since the qubit measurements of the six-state protocol are complete, the input operator $\ket{\psi^+}\bra{\psi^+}$ can be expanded into the
basis $\{F_Q^{(i)} \otimes \sigma_j \}$, where the $\sigma_j$ are the Pauli operators:
\begin{equation*}
  \ket{\psi^+}\bra{\psi^+}=\frac{1}{4} \bigg\{ \mathbbm{1}_Q \otimes \mathbbm{1}_{Q^{\prime}} +
  3\!\!\!\!\!\!\sum_{\alpha=\{x,y,z\}}\!\!\!\! \left( F_Q^{(0,\alpha)} -
    F_Q^{(1,\alpha)}\right) \otimes \sigma^{T}_{\alpha} \bigg\}.
\end{equation*}
This decomposition has the advantage that the adjoint map $\Lambda^\dag$ can be applied directly to the first subsystem by using the substitution $F_Q^{(i)} \mapsto F_M^{(i)}$, which is clear from the properties of the adjoint squasher. This operator $\tau$ has negative eigenvalues, starting in the three photon subspace. For example, if one tests the operator with the state
\begin{equation}
\ket{\theta_{-}}=\frac{1}{\sqrt{2}}\left(\ket{3,0}_{M_{z}}\otimes\ket{1}_{Q'} - \ket{0,3}_{M_{z}}\otimes\ket{0}_{Q'}\right),
\end{equation}
where $\ket{0}_{Q'}$ and $\ket{1}_{Q'}$ are canonical orthogonal basis states, we find $\braket{\theta_{-}|\tau|\theta_{-}}=-1/8$. This proves that a squashing map for the six-state protocol does not exist.

To summarize, we have given necessary and sufficient linear conditions on a positive operator so that a full measurement can be represented by a concatenation of a squashing operation and a lower dimensional target measurement. In application to security proofs of QKD, the existence of a squashing model allows a simple qubit-based security proof to be lifted to one based on the full optical implementation, as is the case for the BB84 measurement, and any other protocol using the same measurement. The squashing model for this BB84 measurement has been independently obtained by Tsurumaru and Tamaki \cite{tsurumaru08suba}. In the absence of a squashing model such a shortcut is not possible and another method of proving security of the full optical scenario has to be found, such as for the six-state measurement. Note that other post-processing methods of the full measurement and target measurements could lead to a squashing model for the six-state protocol detector. As the squashing property holds for the detection set-up independent of the use of the detection device, the method outlined in our Letter will help to simplify the analysis in other quantum communication contexts, including the verification of entanglement of optical modes with threshold detectors.

\begin{acknowledgments}
The authors acknowledge financial support via the European Projects SECOQC and QAP, by the NSERC Discovery Grant, QuantumWorks and CSEC. We also thank Marco Piani and Marcos Curty for helpful discussions and feedback.
\end{acknowledgments}

\bibliographystyle{apsrev}

\end{document}